\documentclass[aps,prl,showpacs,twocolumn,superscriptaddress]{revtex4-1}
\usepackage{amsmath}
\usepackage{amsfonts,bbold}
\usepackage{amssymb}
\usepackage{graphicx}
\usepackage{subfigure}
\usepackage{natbib}
\usepackage[colorlinks=true,linkcolor=blue,urlcolor=black,citecolor=blue]{hyperref}

\begin{document}

\title{Two-component Dirac equation}

\author{Da-Wei Luo}
\affiliation{Beijing Computational Science Research Center, Beijing 100094, China}
\affiliation{Department of Theoretical Physics and History of Science, The Basque Country University (UPV/EHU), PO Box 644, 48080 Bilbao, Spain}
\affiliation{Ikerbasque, Basque Foundation for Science, 48011 Bilbao, Spain}

\author{P. V. Pyshkin}
\affiliation{Beijing Computational Science Research Center, Beijing 100094, China}
\affiliation{Department of Theoretical Physics and History of Science, The Basque Country University (UPV/EHU), PO Box 644, 48080 Bilbao, Spain}
\affiliation{Ikerbasque, Basque Foundation for Science, 48011 Bilbao, Spain}

\author{Ting Yu}
\affiliation{Beijing Computational Science Research Center, Beijing 100094, China}
\affiliation{Center for Controlled Quantum Systems and Department of Physics and Engineering Physics, Stevens Institute of Technology, Hoboken, New Jersey 07030, USA}

\author{Hai-Qing Lin}
\affiliation{Beijing Computational Science Research Center, Beijing 100094, China}

\author{J. Q. You}
\affiliation{Beijing Computational Science Research Center, Beijing 100094, China}

\author{Lian-Ao Wu}
\email{lianaowu@gmail.com}
\affiliation{Department of Theoretical Physics and History of Science, The Basque Country University (UPV/EHU), PO Box 644, 48080 Bilbao, Spain}
\affiliation{Ikerbasque, Basque Foundation for Science, 48011 Bilbao, Spain}

\date{\today}

\begin{abstract}
We provide an alternative approach to relativistic dynamics based on the Feshbach projection technique. Instead of directly studying the Dirac equation, we derive a two-component equation for the upper spinor. This approach allows one to investigate the underlying physics in a different perspective. For particles with small mass such as the neutrino, the leading order equation has a \textit{Hermitian} effective Hamiltonian, implying there is no leakage between the upper and lower spinors. In the weak relativistic regime, the leading order corresponds to a non-Hermitian correction to the Pauli equation, which takes into account the non-zero possibility of finding the lower-spinor state and offers a more precise description.
\end{abstract}


\maketitle

{\it Introduction.}{\bf--}
The Dirac equation~\cite{Dirac1935} offers a quantum mechanical description of the relativistic dynamics of any spin-$1/2$ particles, and is the first theory that merges these two most important discoveries of modern physics. This elegant equation successfully predicts the existence of the antimatter~\cite{Anderson1933}, offers a theoretical justification for the introduction of electron spin and spin-orbit coupling~\cite{Bjorken1964} and the fine-structure of the hydrogen-like atoms~\cite{Bjorken1964}. The Dirac equation also predicts a quivering motion of free relativistic quantum particles called Zitterbewegung~\cite{Schliemann2005,Barut1981,E.-Schrodinger1930}, which is attributed to the interference between the positive and negative energy part of the spinor. Wave packet dynamics of free particles~\cite{Demikhovskii2010} as well as particles in Coulomb potentials~\cite{Arvieu2000,Parker1986} have also been under intense research efforts.

Recently, experimental advances allows for the implementation of various proposals to study the relativistic quantum mechanics phenomena using ion traps~\cite{Gerritsma2010,Lamata2011} as quantum simulators for the Dirac equation, and Zitterbewegung~\cite{Gerritsma2010} as well as the Klein paradox~\cite{Salger2011,Klein1929} have been experimentally observed. This also offers a new approach for other research areas. One example is in Bose-Einstein condensates~\cite{Garay2000}, where the black hole evaporation involves the creation of quasiparticle pairs in positive and negative energy states. Another example is its application in quantum optics, where a mapping between the Jaynes-Cummings model~\cite{Jaynes1963} and the Dirac harmonic oscillator~\cite{Moshinsky1989} is discovered.

While formally simple and elegant, the Dirac equation has some peculiar properties. For example, one needs to change the idea of bare vacuum to an infinite negative energy sea to interpret the negative energy solution for the Dirac equation, which may be quite a hurdle for many. It also employs four components for a relativistic spin-$1/2$ particle, a big departure from the two-component description people are familiar with in the non-relativistic regime. It has been hitherto unclear what a two-component description of the relativistic dynamics would look like or if it is even possible. In this Letter, we ask: can we give a reasonable two-component description for the relativistic dynamics? Indeed, it is often more easy to glean information from the Dirac equation for two-component spinors under some special regime. One interesting regime is for particles with small mass such as neutrinos. Neutrino mass has been experimentally found to be extremely small and theoretically assumed to be zero~\cite{Sakurai1958}. Two-component equations have been suggested ignoring the mass and external field~\cite{Lee1957}. Realistically, it is of great importance to study the different-order contribution of non-zero neutrino mass on the relativistic dynamics of the particle in an electromagnetic field, which has been missing in the literature. On the other hand, the Pauli equation is obtained by a heavily approximated lower spinor in the non-relativistic limit. The Pauli equation provides a good approximation for the gyromagnetic ratio as well as an explanation for the Stern-Gerlach experiment~\cite{Dirac1935,Bjorken1964}. High order correction to the Pauli equation has also been done using the Foldy-Wouthuysen transform~\cite{Foldy1950}, which eliminates the odd terms from the Hamiltonian through a series of canonical transforms. However, a major drawback of this approach is that the effective Hamiltonian in the Pauli equation is Hermitian and produces a unitary evolution for the upper spinor. As a result, for a spin-$1/2$ initially prepared in a state with no lower-spinor component, the Pauli equation predicts that there will be no possibility of finding the lower-spinor, in contradiction to the prediction of the Dirac equation. In this Letter, we provide an alternative approach to solve these issues. By using the Feshbach P-Q partition technique~\cite{Wu2009,Jing2015,Jing2014} for the Dirac equation, we obtain a two-component spinor equation, which may further be cast into a time-convolutionless (TCL) form. Especially, two regimes are investigated, one with small particle mass and the other with weak relativistic effects. It is found that the leading order equation for the small mass case takes on a very compact form and has a Hermitian effective Hamiltonian. In the weak relativistic limit, the leading order equation gives a non-Hermitian correction to the Pauli equation, therefore correctly predict the non-zero possibility of finding the lower-spinor state for an initial state with no lower-spinor component and offers a much more precise perspective.

{\it Feshbach partition for the TCL Dirac equation.}{\bf--}
The Dirac equation merges quantum mechanics with special relativity and has predicted many interesting phenomena, such as spin-orbit coupling. Taking $\hbar=1$ and assuming minimal coupling for the electromagnetic field, we have
\begin{equation}
	i\partial_t\Psi=\left(\beta mc^2+e \varphi +c \pmb{\alpha}\cdot \pmb{\pi}\right)\Psi,\label{eq_dirac}
\end{equation}
where $e$ is the charge carried by the particle, $\pmb{\pi}=\pmb{p}-e\pmb{A}/c$ and $(\varphi, \pmb{A})$ is the vector four-potential for the electromagnetic field. A widely used procedure is to partition the state into upper and lower halves, corresponding to normal particle and lower-spinor solutions with positive energies. It can be very illustrative to study the equation of motion for the upper component. For example, in the non-relativistic approximation, the upper spinor dominates and follows the Pauli equation. Since the effective Hamiltonian of the Pauli equation is Hermitian, the upper spinor evolves unitarily. As a result, this approximation ignores the small but non-zero possibility of finding the negative energy part, i.e., an lower-spinor state. Here we want to derive a time convolution-less equation for the upper spinor by using a systematic projection technique.

To do that, we first use a Feshbach P-Q partition technique~\cite{Wu2009,Jing2015,Jing2014}. Define the projectors
\begin{align}
	\mathcal{P} \equiv\left(
	\begin{array}{c|c}
		\mathbb{1}  & \mathbb{0}  \\ \hline
		\mathbb{0} &  \mathbb{0}\\
	\end{array}\right), \;
	\mathcal{Q} \equiv \mathcal{I}-\mathcal{P} =\left(
	\begin{array}{c|c}
		\mathbb{0}  & \mathbb{0}  \\ \hline
		\mathbb{0} &  \mathbb{1}\\
	\end{array}\right),
\label{FQ-proj}
\end{align}
where $\mathbb{0}$ and $\mathbb{1}$ are both $2 \times 2$ matrices. the wave function $\Psi =[\Psi_1,\Psi_2,\Psi_3,\Psi_4]^T$ can the be partitioned as $\mathcal{P}\Psi =[\Psi_1,\Psi_2,0,0]$ and $\mathcal{Q}\Psi =[0,0,\Psi_3,\Psi_4]^T$, where $T$ stands for matrix transpose. Accordingly, the Hamiltonian can be partitioned into $4$ two-by-two matrices as
 \begin{equation}
	H=\left(
	\begin{array}{c|c}
	\tilde{h}&\tilde{R}\\ \hline
	\tilde{W}&\tilde{D}
	\end{array}
	\right),
\end{equation}
where $\tilde{h}$, $\tilde{R}$, $\tilde{W}$, $\tilde{D}$ are non-zero matrix blocks corresponding to $h=\mathcal{P}H \mathcal{P}$, $R=\mathcal{P}H \mathcal{Q}$, $W=\mathcal{Q}H \mathcal{P}$ and $D=\mathcal{Q}H \mathcal{Q}$. The exact integral-differential equation for the upper spinor is then given by
\begin{align}
	&i \partial_t \mathcal{P}|\psi (t)\rangle=e \varphi \mathcal{P}|\psi (t)\rangle \nonumber\\
	&-i c^2 \int_0^tds \left\{\phi(t-s)\pmb{\sigma}\cdot \left[-eA/c-e(t-s) \nabla \varphi \right]\right\}\pmb{\sigma}\cdot \pmb{\pi}\mathcal{P}|\psi (s)\rangle \nonumber\\
	&-i c^2 \int_0^tds \phi(t-s) \left[\pmb{\pi}^2-e \pmb{\sigma}\cdot \pmb{B}/c\right]\mathcal{P}|\psi (s)\rangle,
\end{align}
where $\phi(t-s)=\exp[i(2mc^2-e \varphi)(t-s)]$. Depending on the problem under consideration, we take the dominant part of the Hamiltonian as $H_0$ and work in the interaction picture with respect to it, i.e., $i\dot{\psi}=H_I(t)\psi$, where $H_I(t)=U_0 ^\dagger(t)(H-H_0)U_0(t)$, $\psi=U_0 ^\dagger(t)\Psi$, and $U_0(t)$ is the propagator associated with $H_0$. Applying the P-Q partition, 
ans assuming we start with a particle state,
we can formally solve for $\mathcal{Q}\psi(t)$ and get
\begin{equation}
	\partial_t \mathcal{P}\psi(t)=-i \mathcal{P}H_I(t) \mathcal{P}\psi(t)-\int_0^t ds \mathcal{C}(t,s) \mathcal{P}\psi(s),\label{eq_ex_nz}
\end{equation}
where $\mathcal{C}(t,s)=\mathcal{P}H_I(t)v(t,s)\mathcal{Q}H_I(s)$ is the memory kernel, $v(t,s)=\widehat{T}\{\exp[-i\int_s^t \mathcal{Q}H_I(\tau)d\tau]\}$ and $\widehat{T}$ is the time-ordering operator. This is the exact Nakajima-Zwanzig equation for the state vector $\mathcal{P}\psi$.

We now cast the equation into a time-convolutionless form by using a time local projection~\cite{Breuer2002}. Writing the formal solution for $\mathcal{Q}\psi(t)$ as $[1-\Sigma(t)]\mathcal{Q}\psi(t)=\Sigma(t)\mathcal{P}\psi(t)$, where $\Sigma(t)=-i\int_0^tdsv(t,s) \mathcal{Q}H_I(s) \mathcal{P}u ^\dagger(t,s)$ and $u(t,s)=\hat{T}\exp[-i\int_s^t d\tau H_I(\tau)]$, we get
\begin{equation}
	\partial_t \mathcal{P}\psi(t)=\mathcal{K}(t)\mathcal{P}\psi(t),
\end{equation}
where $\mathcal{K}=-i \left\{\mathcal{P}H_I(t) \mathcal{P}+\mathcal{P}H_I(t)[1-\Sigma(t)]^{-1}\Sigma(t)\mathcal{P}\right\}$ is the TCL generator. The invertibility of the operator $1-\Sigma(t)$ is ensured due to the fact that it is a perturbation of the identity operator since $\lim_{H_I\rightarrow 0}\Sigma(t)=0$. We can now expand $[1-\Sigma(t)]^{-1}\Sigma(t)=\sum_{k=1}\Sigma^k(t)$, up to any order of $H_I$. 

As a first application, we consider a particle with very small mass in a static field, such as the neutrino particle. In this case, $H_0=e \varphi +c \pmb{\alpha}\cdot \pmb{\pi}$, and $H_I(t)=m h_I(t)$, where $h_I(t)$ is mass independent. At the leading order of mass $m$, we have 
\[
\partial_t \mathcal{P}\psi(t)=-im \mathcal{P}h_I(t)\mathcal{P}\psi(t),
\]
which, remarkably, has a \textit{Hermitian} effective Hamiltonian, generating a unitary propagator. Therefore, for a state initially prepared in the $\mathcal{P}$-space, i.e., $\mathcal{Q}\psi(0)=0$, it will stay in the $\mathcal{P}$-space up to the first order. Especially, in absence of external field, we explicitly have $\partial_t \mathcal{P}\psi(t)=-2imc^2\cos^2(c|\pmb{p}|t)\mathcal{P}\psi(t)$ as a first-order approximate equation, where $|\pmb{p}|$ denotes the norm of the momentum $\pmb{p}$. The equation has a plane wave solution,
\[
\int dp c_p(0)e^{ipx-imc^2t[1+\mathrm{sinc}(2cpt)]},
\] 
where $c_p(0)$ is the initial condition. Up to $O(m^2)$, this is in agreement with the plane wave solution obtained by directly solving the Dirac equation.

On the other hand, in the weak relativistic regime, we have a dominant diagonal Hamiltonian which we take as $H_0$. At the leading order, we have $\partial_t \mathcal{P}\psi(t)=-\left[\int_0^t ds \mathcal{P} H_I(t)H_I(s) \mathcal{P}\right]\mathcal{P}\psi(t)$. Going back to the original picture
and rotating out a trivial global phase $\exp \left[-imc^2t\right]$ for the whole Hamiltonian, we have
\begin{align}
	\partial_t& \mathcal{P}\Psi (t)=\left[-ie \varphi-c^2\int_0^tds \left(\pmb{\sigma}\cdot \pmb{\pi} \right) \exp\left[-ie \varphi (t-s)\right]\right.  \nonumber\\
	&\left.\left(\pmb{\sigma}\cdot \pmb{\pi}\right)\exp\left[ie \varphi (t-s)\right] \exp\left[2imc^2 (t-s)\right] \vphantom{\int_i^f}  \right] \mathcal{P}\Psi (t).
\end{align}
Using $\left[\pmb{A},\varphi \right]=0$, $[\pmb{p},f(\pmb{q})]=-i \nabla f(\pmb{q})$ and the BCH formula $\exp[{A}] B \exp[{-A}]=B+\sum_{m=1}^{\infty}[_m A,B]/m!$, where $[_m A,B]=[A,[_{m-1} A,B]]$ and $[_1 A,B]=[A,B]$, we can simplify the equation and arrive at
\begin{widetext}
\begin{align}
	\partial_t& \mathcal{P}\Psi (t)=-i\left[e \varphi+ \left(\frac{\pmb{\pi}^2}{2m}-\frac{e \pmb{\sigma}\cdot \pmb{B}}{2mc}\right)\left(1-\exp(2imc^2 t)\right)\right]\mathcal{P}\Psi (t) \nonumber\\
	& -i\left[\frac{e} {4m^2c^2} \left[ \left(\nabla \cdot \nabla \varphi\right)+i(\nabla \varphi)\cdot\pmb{\pi}+\pmb{\sigma}\cdot (\nabla \varphi)\times \pmb{\pi} \right] \left[1-\exp(2imc^2 t)(1-2imc^2t)\right]  \right] \mathcal{P}\Psi (t).\label{eq_tcld}
\end{align}
\end{widetext}
We recognize the first line of the equation as a non-Hermitian correction to the Pauli equation with an effective Hamiltonian $H_{\rm eff}=e \varphi+\pmb{\pi}^2/2m-e \pmb{\sigma}\cdot \pmb{B}/2mc$ since the long time average of $\exp(2imc^2 t)=0$. The second line is of order $e/4m^2c^2$ and is therefore a higher order correction. The effective Hamiltonian TCL equation is no longer Hermitian, and tracks the non-zero possibility of finding the lower-spinor state up to the leading order. Higher order equation can be obtained in the same fashion by including higher order of $\Sigma^k(t)$.

\begin{figure}[t!]
	\centering
	\subfigure[]{%
	\includegraphics[scale=.55]{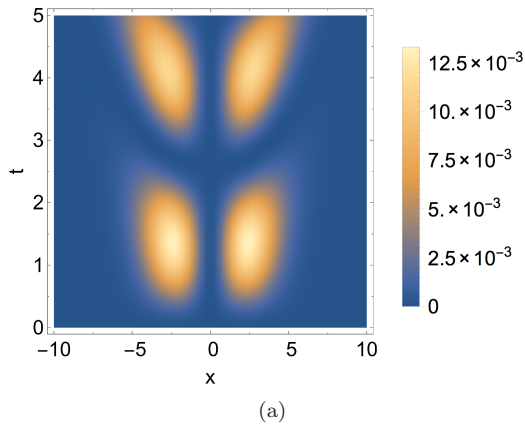}
	\label{fig1a}}
	\subfigure[]{%
	\includegraphics[scale=.55]{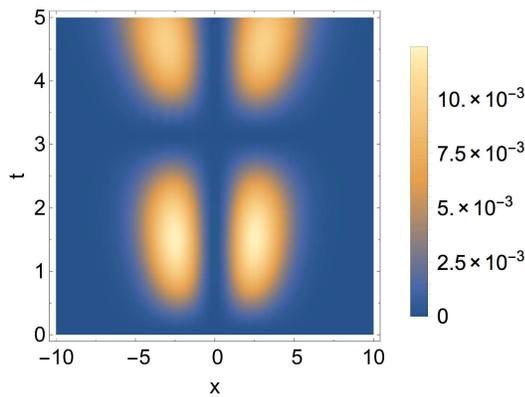}
	\label{fig1b}}
	\caption{Density plot for the possibility of finding lower-spinor state as a function of position $x$ and time $t$ with no electromagnetic field, $m=c=1$, $x_0=10$ and $p_0=0$. Panel (a) is obtained from the Dirac equation and panel (b) is obtained from the TCL equation, Eqs.~\eqref{eq_tcld} and~\eqref{eq_tcld_q}. A good agreement between the two can be observed.}
	\label{fig1}
\end{figure}

\begin{figure}[t!]
	\centering
	\subfigure[]{%
	\includegraphics[scale=.55]{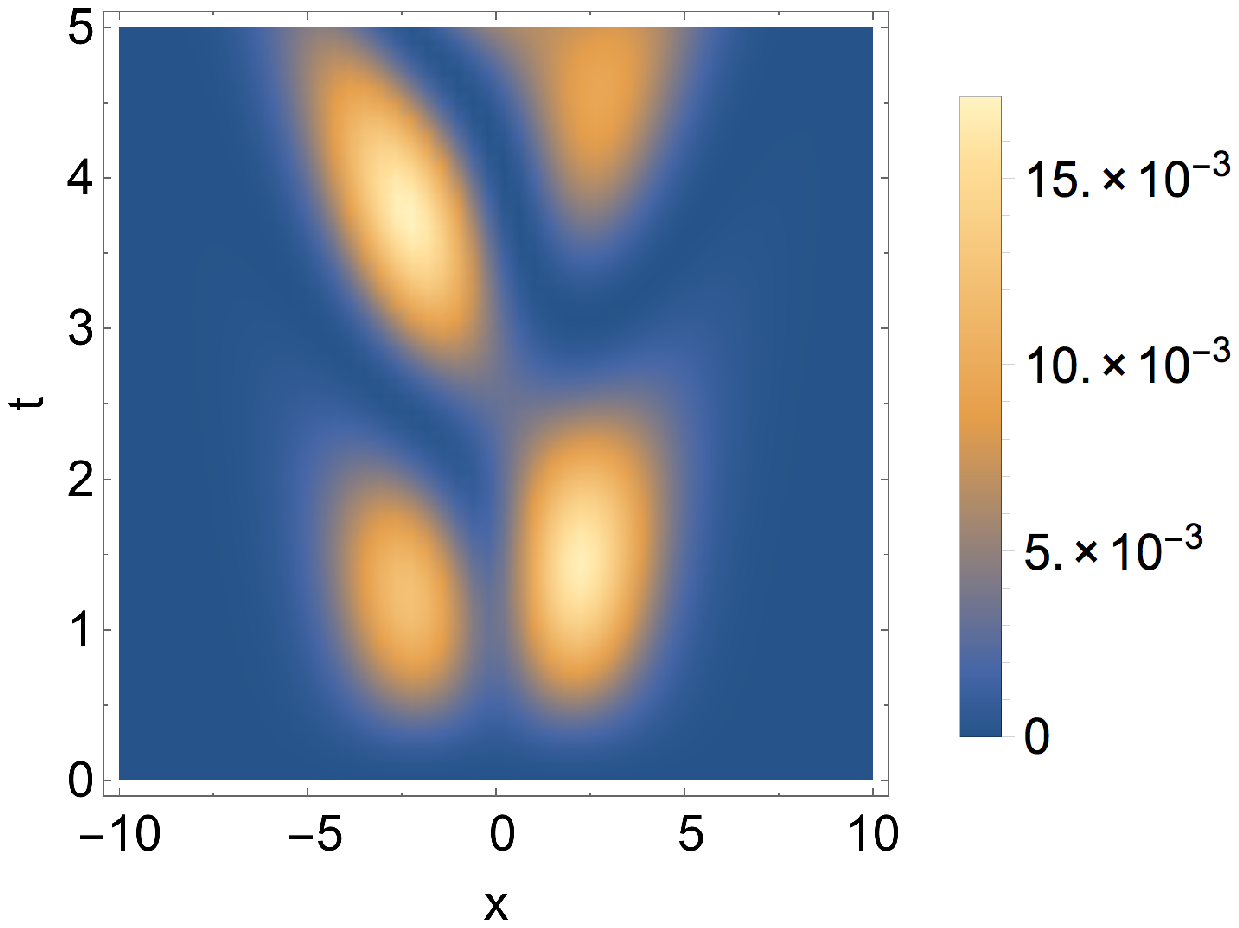}
	\label{fig2a}}
	\subfigure[]{%
	\includegraphics[scale=.55]{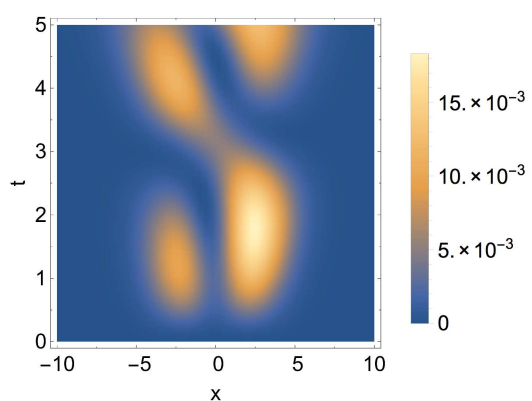}
	\label{fig2b}}
	\caption{Density plot for the possibility of finding lower-spinor state as a function of position $x$ and time $t$ with a static potential $e\varphi=ax$. Parameters used are $m=c=1$, $a=0.1$, $x_0=10$ and $p_0=0.2$. Panel (a) is obtained from the Dirac equation and panel (b) is obtained from the TCL equation, which is shown to offer a good approximation of the exact dynamics.}
	\label{fig2}
\end{figure}

{\it Examples.}{\bf--} As an illustrative example, we first consider a free relativistic particle, under zero electromagnetic field. The Dirac equation (Eq.~\eqref{eq_dirac}) and the TCL equation (Eq.~\eqref{eq_tcld}) are analytically solvable as planar waves. We choose a Gaussian wave packet for the upper spinor as $f(x)=\sqrt[4]{2/x_0 \pi}\exp[-x^2/x_0+ip_0x]$, corresponding to a Gaussian wave packet centered around $p_0$ in the momentum space. The lower spinor is initially set to zero. Therefore, any non-zero $\mathcal{Q}\Psi(\pmb{q},t)$ means a non-zero probability of finding the lower-spinor at position $\pmb{q}$, which is ignored by the Pauli equation. We can use $1-\int d\pmb{q}\mathcal{P}\Psi(\pmb{q},t)$ to quantify the total possibility of finding the lower-spinor at all positions at time $t$, but a more intricate formula including the positional dependence can be used. To get that, we use the corresponding $\mathcal{Q}$ part of Eq.~\eqref{eq_tcld}, $\mathcal{Q}\psi(t)\approx -i\left[\int ds \mathcal{Q}H_I(s) \mathcal{P} \right]\mathcal{P}\psi(t)$. Going back to the original picture, we have
\begin{align}
	\mathcal{Q}&\Psi (t)={-i}\left[\left(2imc^2 \pmb{\sigma}\cdot\pmb{\pi}-e\pmb{\sigma}\cdot \nabla \varphi \right)\left(1-\exp[{2imc^2t}]\right)\right. \nonumber\\
	&\left.-2imc^2t \exp[{2imc^2t}]e\pmb{\sigma}\cdot \nabla \varphi \right] \mathcal{P}\Psi (t)/\left({4m^2c^3}\right).~\label{eq_tcld_q}
\end{align}
For simplicity, we can study the 1D equation without loss of generality. In this case, the upper and lower spinor can be described by 1 component each, and the eigenvector of the Dirac Hamiltonian is
\begin{align*}
	&U_+=\sqrt{\frac{\lambda+mc^2}{2 \lambda}}\left(\begin{array}{c}1\\pc/(\lambda+mc^2)\end{array}\right), \nonumber \\
	&U_-=\sqrt{\frac{\lambda+mc^2}{2 \lambda}}\left(\begin{array}{c}-pc/(\lambda+mc^2)\\1\end{array}\right),
\end{align*}
with eigenvalues $\pm \lambda$, where $\lambda=\sqrt{p^2c^2+m^2c^4}$. The solution of the TCL equation is given by
\begin{align*}
	\mathcal{P}\Psi&=\int dp \exp[ipx] c_p(0) \\
	&\times \exp \left[\frac{p^2}{4m^2c^2}\left[\exp(2imc^2t)-2imc^2t-1\right]\right],
\end{align*}
where $c_p(0)$ is determined by a Fourier transform of the initial state in the position space.

In Fig.~\ref{fig1} we plot $|\mathcal{Q} \Psi(x,t)|^2$ as a function of position $x$ and time $t$ using the exact solution via the Dirac equation in panel (a) and via the TCL equation in panel (b), choosing $m=c=1$, $x_0=10$ and $p_0=0$. It can be observed that the TCL equation can approximate the non-zero probability of finding the lower-spinor predicted by the Dirac equation, a fact that's totally ignored in the Pauli equation.

As a second example, we choose a linear static linear field $e\varphi=ax$ and numerically solve the exact equation and the TCL equation. Choosing $m=c=1$, $a=0.1$, $x_0=10$ and $p_0=0.2$, $|\mathcal{Q} \Psi(x,t)|^2$ as a function of position $x$ and time $t$ is shown in Fig.~\ref{fig2}, where panel (a) is obtained using the exact Dirac equation and panel (b) is obtained via the TCL equations, where a good agreement between the two is also observed. Therefore, the TCL equation can give us a more precise two-component description for the relativistic particle than the Pauli equation.

{\it Conclusion.}{\bf--} In conclusion, by using a Feshbach P-Q partition and a time-local projection with the Dirac equation, we obtain a two-component equation for the upper spinor, which can be further be cast into a TCL form. This alternative approach allows for a different perspective to study the relativistic dynamics for spin-$1/2$ particles. Both the small mass regime and the weak relativistic regimes are investigated. The leading order equation in the small mass regime takes a compact form. Remarkably, the effective Hamiltonian for the upper spinor is Hermitian at the leading order, predicting that the particle will stay in the $\mathcal{P}$ space as a first order approximation. For the weak relativistic regime, unlike the Pauli equation whose effective Hamiltonian for the upper spinor is Hermitian, the TCL equation obtained here is non-Hermitian and correctly takes into account the non-zero probability of finding the lower-spinor state.

{\it Acknowledgments.}{\bf--}
This work is supported by the Basque Government (Grant No.~IT472-10), the MINECO (Project No.~FIS2012-36673-C03-03), and the Basque Country University UFI (Project No.~11/55-01-2013). J. Q. You is supported by the National Natural Science Foundation of China No.~91421102 and the National Basic Research Program of China No.~2014CB921401. T.Y. is supported by the NSF PHY-0925174 and DOD/AF/AFOSR No.~FA9550-12-1-0001.




\end{document}